# Linking Properties to Microstructure in Liquid Metal Embedded Elastomers via Machine Learning

Abhijith Thoopul Anantharanga[1], Mohammad Saber Hashemi[1], Azadeh Sheidaei[1*]

[1] Aerospace Engineering Department, Iowa State University, Ames, IA 50011, United States

**Abstract**

Liquid metals (LM) are embedded in an elastomer matrix to obtain soft composites with unique thermal, dielectric, and mechanical properties. They have applications in soft robotics, biomedical engineering, and wearable electronics. By linking the structure to the properties of these materials, it is possible to perform material design rationally. Liquid-metal embedded elastomers (LMEEs) have been designed for targeted electro-thermo-mechanical properties by semi-supervised learning of structure-property (SP) links in a variational autoencoder network (VAE). The design parameters are the microstructural descriptors that are physically meaningful and have affine relationships with the synthetization of the studied particulate composite. The machine learning (ML) model is trained on a generated dataset of microstructural descriptors with their multifunctional property quantities as their labels. Sobol sequence is used for in-silico Design of Experiment (DoE) by sampling the design space to generate a comprehensive dataset of 3D microstructure realizations via a packing algorithm. The mechanical responses of the generated microstructures are simulated using a previously developed Finite Element (FE) model, considering the surface tension induced by LM inclusions, while the linear thermal and dielectric constants are homogenized with the help of our in-house Fast Fourier Transform (FFT) package. Following the training by minimization of an appropriate loss function, the VAE encoder acts as the surrogate of numerical solvers of the multifunctional homogenizations, and its decoder is used for the material design. Our results indicate the satisfactory performance of the surrogate model and the inverse calculator with respect to high-fidelity numerical simulations validated with LMEE experimental results.

Corresponding author: Azadeh Sheidaei, PhD, Aerospace Engineering Department, Iowa State University, Ames, IA 50011, United States, Tel: 515-294-2956 (O)/Fax: 515-294-3262 (O), Email: Sheidaei@iastate.edu

**Keywords:** Liquid metal embedded elastomers, Multifunctional properties, Flexible composites, Computational material design, Machine learning

1. **Introduction**

The main goal of computational material design is to perform inverse design so that novel materials with unique properties are discovered. Therefore, process-structure-property linkage should be established to develop a methodology for inverse design of materials and to understand physics behind microstructure generation [1–6]. The main idea is to create a large dataset of microstructural properties and explore this dataset via neural networks to obtain insights which are used in the discovery of novel materials with unique, desired properties. In recent years, there has been a shift from unifunctional materials to multifunctional materials. Hence, computational material design as a burgeoning interdisciplinary field of study has been developed to design multifunctional materials with target properties. The first step for such studies is to computationally characterize and reconstruct the microstructures as they are determinant of the target properties. Physical descriptors are used in this study based on the specific morphological characteristics of particulate composite microstructures studied, their close relationship to the parameters used and measured in synthetization, their clear physical interpretation, and low dimensionality [7]. ML-based methods have been used to relate physical descriptors to target variables and to model the complex relationships between them [7–9].

Interest in particulate composites having tiny inclusions embedded in a matrix has increased since they can be designed and manufactured to improve on a single property or multiple properties of constituents in their single-phase formats and to achieve novel physical responses or functionalities. For example, inorganic fillers are added to soft polymers to obtain composites with a combination of mechanical properties from polymers and electrical and thermal properties from inorganic fillers [10–12]. Rubber is added to silicone elastomers to obtain composites with high dielectric constant [13]. The material properties depend on the size, geometry, distribution, and aspect ratio of the inclusion materials[14]. LMEE composites as a class of multifunctional particulate composite materials with simultaneously tunable mechanical, electrical, and thermal properties are the focus of our study. They take advantage of unique combination of high thermal conductivity and high dielectric constant of Liquid Metal (LM) inclusions and high stretchability and fracture toughness of the elastomer matrix [15]. They have applications in robotics, stretchable electronics, and biomedical engineering due to their capabilities in storage of energy, actuation, stretching and sensing [15–19]. Over the past few years, ways to synthesize these LM droplets in matrix materials have been explored. Until now, LM inclusions of Eutectic Ga-In are synthesized with polydisperse suspensions of micro-sized droplets in the matrix [15]. With more such developments in synthesizing these materials such as the recent work of Haake et al. [20], a computational framework to design LMEEs with target properties is needed. Dehnavi et. al. developed an FFT micromechanical computational methodology for quick and efficient analysis of the effective thermal properties of particulate composites [21]. Chiew and Malakooti [22] developed a double inclusion model capable of predicting properties of polymer composites with core-shell liquid metal droplets. But this model is applicable for spherical inclusions and considers linear elastic behaviour of constituents. Also, it does not consider the distribution of particles of various sizes in the microstructure since it is a micromechanical model. Later Hashemi et. al. utilized ML approaches to predict thermal conductivities in particulate composites [7].

For direct surrogate SP modeling and inverse SP design calculation, such ML-based frameworks in their inference mode are faster and cheaper than brute-force repetition of computationally intensive numerical simulations as well as the Edisonian trial and error in numerous physical experiments [1], and they can be more accurate than mean-field continuum models which are based on some simplifying assumptions. To the best of authors' knowledge, there has been no study on the robust multifunctional and computational SP relationships of LMEEs and other similar material systems which is necessary to design them for their multifunctional applications with target mechanical, dielectric, and thermal properties. In this study, a framework that targets mechanical, thermal, and dielectric properties has been developed. This framework not only predicts these properties but also helps understand the complex relationship between the material microstructure and its properties via a semi-supervised variational autoencoder trained on an in-silico and sufficiently big dataset of microstructures represented by their interpretable and physical microstructural descriptors as well as their computationally homogenized properties. The graphical abstract of the computational framework is shown in Fig. 1. In Section 2, the methodology is explained. Section 2.1 elaborates on how microstructure is realized to generate the training dataset. Section 2.2 explains the numerical simulations, namely, FEA and FFT. Section 2.3 covers the ML models used in SP links. Section 3 includes the results and discussions. Section 3.1 elaborates on why our ML model is preferred over analytical models. Section 3.2 explains the performance of the surrogate model and the inverse designer. Section 3.3 explains the relationship between the material properties and microstructural physical descriptors. Finally, Section 4 includes the conclusions that can be drawn from this study.

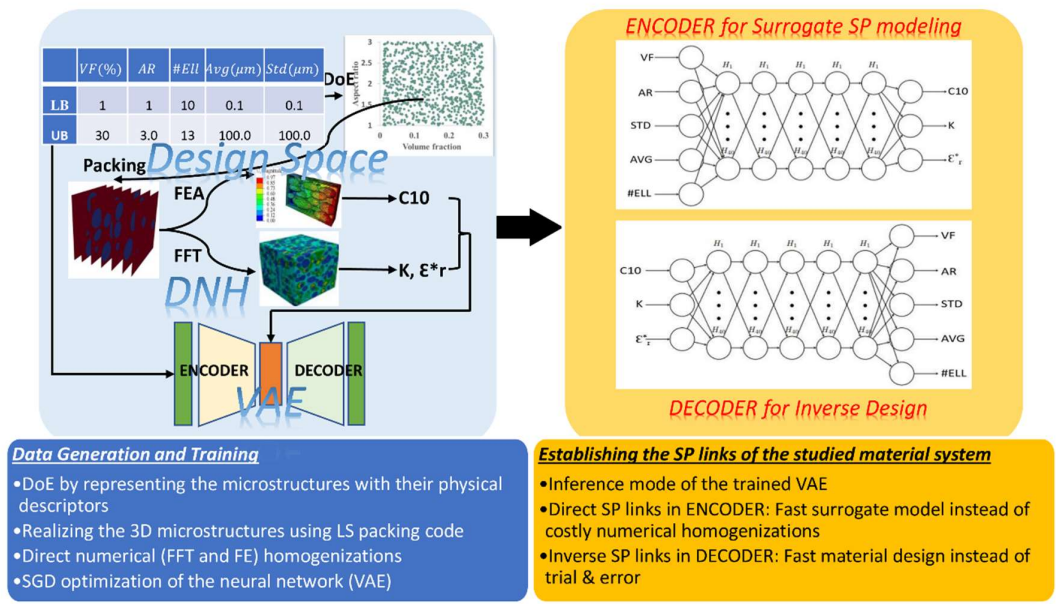

**Fig. 1**. Overview of our ML-based computational framework for designing the particulate composites

## 2. Materials and Methods

*2.1. Microstructures' realization to generate the training dataset*

The microstructures are characterized by their physical microstructural descriptors, i.e., the volume fraction of the LM inclusions, aspect ratio of ellipsoidal inclusions, and the mean and standard deviation of their sizes. The volume fraction of LM inclusions has been restricted to 30% due to cost and time prohibition of packing over this volume fraction. LM inclusions can be either spherical or ellipsoidal in shape, so an equal number of spheres were assumed to be inside each microstructure to better simulate the real material microstructure. The number of inclusions is made dependent on the volume fraction and is determined as ($volume\_fraction \times 10 + 10$), rounded off to nearest whole number for both spherical and ellipsoidal inclusions. The descriptors' bounds in Table 1 are determined based on the experimental studies [18] on isotropic LMEEs, and they delineate the hypercube of material design space. The descriptors are considered as design variable dimensions of a 4D quasi-random Sobol sequence [23] in order to generate a sufficiently large and uniformly distributed training dataset due to Sobol's low-discrepancy properties as done in [7]. Each data entry, a numerical 4D vector, describes the characteristics of an LMEE microstructure, which is then geometrically realized via Lubachevsky-Stillinger (LS) algorithm [24] for packing disks and ellipsoids such that they do not overlap with each other and do not touch each other. The packing boundary conditions were considered as rigid or solid walls as opposed to the periodic alternative so that there were no partially open LM inclusions not suitable for the FE mechanical simulations. To further ensure the uniformity of the training dataset and due to the a priori knowledge of larger sensitivity of properties to VF changes [7,15,18], packs were divided into 20 bins with each bin covering a small range of the volume fraction, and 50 packs in each bin were selected, totaling 1000 packs. A single input of the packing code is the processed numerical vector of the Sobol sequence, and its raw output is the numerical information of the number of inclusions, the coordinates of their centers, their sizes, their aspect ratios, the mean and standard deviation of the inclusion sizes. Since FE and FFT methods operate on the discretized mesh of elements and the 3D or voxelized image of the microstructure, respectively, and FE is more computationally demanding requiring the minimum number of elements conforming to the real geometry, two geometrical representations were obtained for each pack via Abaqus scripting and the packing code's meshing subroutine, respectively. Each generated 3D realization has its dimensions normalized to 1×1×1 and consists of 300 2D sliced images of each pack in the voxelized representation ($300^3$ voxels) as it is shown in Fig. 2.

The microstructures generated in this study have a domain size calculated like in [7] where it was proved that the generated microstructures were in fact RVEs by plotting the thermal conductivity against the number of particles and showing that it is almost constant. These domain sizes were seen to be dependent on the physical descriptors of the microstructures. Since the same methodology is adopted in this study to obtain the domain size, the microstructures in this study are RVEs.

**Table 1:** The bounds on the physical microstructural descriptors controlled

|  | $VF(\%)$ | $AR$ | $\#Ell$ | $Avg(\mu m)$ | $Std(\mu m)$ |
|---|---|---|---|---|---|
| Lower Bounds | 1 | 1 | 10 | 0.1 | 0.1 |
| Upper Bounds | 30 | 3.0 | 13 | 100.0 | 100.0 |

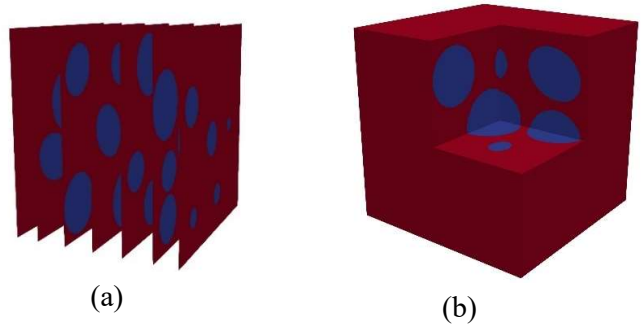

(a)           (b)

**Fig. 2.** (a) the stack of 2D images of a 3D voxelized model, (b) the 3D visualization of the pack

*2.2. Quantification of properties through direct numerical homogenization*

The target properties are the composite hyperelastic constant as well as its effective thermal conductivity and dielectric constants. These properties are calculated using homogenization techniques. The mechanical properties of the LMEEs are obtained via FE simulations and optimizations. The matrix material is a silicone gel with an initial shear modulus $\mu_0$ of $566.67\ Pa$ and an initial bulk modulus $K_0$ of $56.67\ kPa$. The inclusion material is Eutectic Gallium-Indium (EGaIn) with a density of $6250\ \frac{kg}{m^3}$ [25]. The surface tension of the matrix-inclusion interface is $0.6\ \frac{N}{m}$ [11]. The matrix as well as the composite material is nearly incompressible and modeled as hyperelastic materials following the Neo-Hookean law. Therefore, the homogenization problem boils down to determining the constant(s) of the effective hyperelastic energy form of the heterogenous material such that the mechanical response of the homogenized model closely resembles that of the real heterogenous one. The responses are the volume averaged stress and strain components throughout the whole model (all elements). To obtain macroscopic stresses and strains, FE simulations were conducted on each microstructure using an Abaqus UEL developed by Yuhao Wang et al. [26] for soft solids with liquid inclusions in which inclusion are neglected in the FE model by considering their surface tension interaction with the matrix. Since the studied material is isotropic, a single tensile test is simulated by applying boundary conditions and a stretch of 90%. The Neo-Hookean strain energy can be expressed as $U = C_{10}(\bar{I}_1 - 3) + \frac{(J^{el}-1)^2}{D_1}$, where $C_{10}$, the hyperelastic constant adjusting the effect of the first deviatoric strain invariant $\bar{I}_1$, and $D_1$, the hyperelastic constant adjusting the effect of to the elastic volume ratio or the Jacobian of the deformation gradient tensor $J^{el}$, are calculated as $C_{10} = \frac{\mu_0}{2}$ and $D_1 = \frac{2}{K_0}$. To impose the near incompressibility condition on the homogenized model in the implicit step solver of ABAQUS, $D_1$ is assumed to be dependent on $C_{10}$ such that the initial Poisson's ratio is greater than 0.495, or $\frac{\mu_0}{K_0} = 1000$. Our previously developed algorithm has been used to compute the effective hyperelastic parameter of the composites [27]. The homogenized stress values of the heterogeneous model at different loading steps are the optimization targets. A homogeneous model consisting of a single material is also modeled with the same boundary conditions, while the homogenous model's $C_{10}$ is the input of an objective function which is the integral of differences between the average stresses of the homogeneous and heterogeneous models at multiple loading steps (different strains). The optimization algorithm minimizes the function by changing the input and getting the associated mechanical responses from the FE solver.

Fig. 3(a) shows one composite microstructure before and after the application of the tensile load as well as its inclusions elongated in the direction of the applied load which makes the whole composite orthotropic. This closely resembles the experimental results of Bartlett et. al. creating a network of elongated LM inclusion by stretching the composite and freezing it to tune the thermal conductivity in that direction [10], thereby the deformed models can be readily used for future studies on anisotropic material design. Fig. 3(b) and (c) are the contour plots of component 11 of the stress tensor, and maximum pricipal logarithmic strain, respectively, where they peak in regions close to inclusions. It can be seen from Fig. 3(b) and Fig. 3(c) that the inclusions are more stressed at the center and less at the ends.

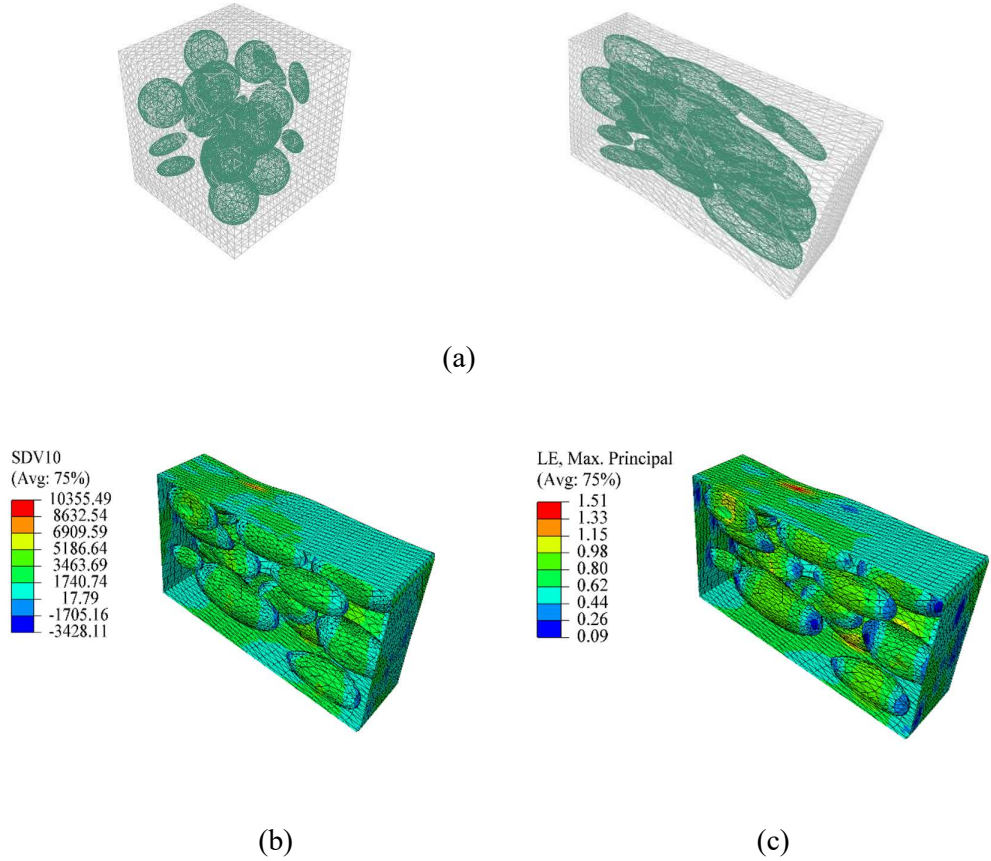

**Fig. 3.** a) An example of the studied composite before and after loading; b) the contour plot of Stress$_{11}$ (Pa) in the FE model, and (c) the contour plot of the logarithmic strain tensor.

Thermal and dielectric properties of the LMEEs are determined using our recently developed Fast Fourier Transform (FFT) computational scheme [21]. The thermal conductivity values of the silicone matrix and EGaIn inclusions are 0.29 W/mK and 26.4 W/mK [21], respectively, and the dielectric constants are 4.9 [28] and 40 [29], respectively. These values were utilized in calculating the thermal conductivity and dielectric constant for LMEEs.

The constitutive equation for thermal conductivity is given as $\bar{q} = K\bar{e}$, where $K$ is the thermal conductivity, $\bar{e}$ is the temperature gradient vector, and $\bar{q}$ is the heat flux vector. The thermal conductivity, $K$ is denoted by $K = K_{ij}; i,j = 1,2,3$. Since the material studied is isotropic, the mean of the diagonal terms or $K = \frac{K_{11} + K_{22} + K_{33}}{3}$ was considered. The constitutive equation for permittivity is given by $\overline{\mathbf{D}} = \mathcal{E}\overline{\mathbf{E}}$, where $\mathcal{E}$ is the absolute permittivity of material, $\overline{\mathbf{D}}$ is the electric displacement field, and $\overline{\mathbf{E}}$ is the applied electric field. The dielectric constant is given by $\mathcal{E}_r^* = \frac{\mathcal{E}}{\mathcal{E}_0}$ with $\mathcal{E}_0$ representing the vacuum permittivity. The dielectric constant is denoted by $\mathcal{E}_r^* = \mathcal{E}_{r_{ij}}^*; i,j = 1,2,3$. Due to material isotropy, it was obtained as $\mathcal{E}_r^* = \frac{\mathcal{E}_{r_{11}}^* + \mathcal{E}_{r_{22}}^* + \mathcal{E}_{r_{33}}^*}{3}$.

*2.3. ML model for SP links*

The ML model is used to robustly link the material microstructure to its properties directly and inversely. The former provides us with a fast (or almost instantaneous after training) surrogate model to predict the material properties without resorting to computationally demanding numerical homogenizers, while the latter creates a fast design calculator for experimentalists interested in composites with multifunctional target properties, which is much needed and hard to come by through other optimization methods. Fully connected neural networks were used since the structure and properties are low dimensional vectors, and the general approximation capability [30] of neural networks is preferable for learning the nonlinear and complex SP relationships. The natural architecture choice is a generative type of neural networks such as a VAE whose encoder maps the distribution of the physical microstructural descriptors (number of inclusions, mean and standard deviation of the inclusion size, aspect ratio of inclusions and volume fraction of the LM phase) to the latent distribution of property vectors (thermal conductivity, hyperelastic parameter and the dielectric constant); whose decoder decodes the sampled vectors of the latent space into vectors of a distribution similar to the first one (microstructural descriptors). VAEs were first introduced by Kingma et al. [31] and have been developed and adopted in different tasks such as generating handwritten digits [31] and numbers [32]. Generally, KL divergence is considered as their loss function to minimize the difference between the decoder's output distribution and encoder's input one. However, in this study, the loss function is the MSE loss function. This is because it was seen that the MSE loss function provides better results compared to the KL divergence. This loss function led to good performances on the test dataset as shown in the results section.

$Loss = L_{encoder} + L_{decoder}$, where, $L_{decoder} = L1 = \frac{1}{N}\sum_{i=1}^{N}(\bar{x}_\iota - \bar{\hat{x}}_\iota)^2$, $L_{encoder} = L2 = \frac{1}{N}\sum_{i=1}^{N}(\bar{y}_\iota - \bar{\hat{y}}_\iota)^2$, where $\bar{x}_\iota$ is the true vector of physical descriptors, and $\bar{\hat{x}}_\iota$ is the predicted vector of physical descriptors; $\bar{y}_\iota$ is the true vector of the material properties, and $\bar{\hat{y}}_\iota$ is the predicted vector of material properties. Here, $\bar{x}_\iota, \bar{\hat{x}}_\iota \in \mathbb{R}^5$ and $\bar{y}_\iota, \bar{\hat{y}}_\iota \in \mathbb{R}^3$. Several combinations of different number of layers, different hyper parameters, and different optimizers were grid-searched to find the best model and training parameters. The dataset was normalized to stabilize the performance in the training process and to achieve better test performances in the inference. Following the grid-search, ReLU and Adams optimizer with a learning rate of 0.001 were chosen as the activation function type and the training optimizer in PyTorch [33], respectively. The best architecture, as shown in Fig. 4, was chosen based on the average performances in a 5-fold cross-validation scheme. The encoder and the decoder consist of 5 fully connected perceptron layers having 40 neurons per each. The input layer has 5 and 3 neurons in the encoder and the decoder, respectively, while the output layer has 3 and 5 neurons, respectively. The network architecture has been further explained in Table 2. The model accuracy was calculated for each material property and is reported in the results section. Three different metrics were used to assess the accuracy. Normalized RMSE is calculated as $NRMSE = \frac{1}{(y_{max} - y_{min})}\sqrt{\frac{1}{N}\sum_{1}^{N}(y_i - \hat{y}_i)^2}$. The mean maximum error is calculated as $MME = \frac{1}{N}\sum_{1}^{N}(y_i^{max} - \hat{y}_i^{max})$, where $y_i$ is the true value of the material property, $\hat{y}_i$ is the predicted value of the material property and N is the number of data entries in the test set. The coefficients of determination ($R^2$) of the regression plots were calculated using the scikit learn library in python.

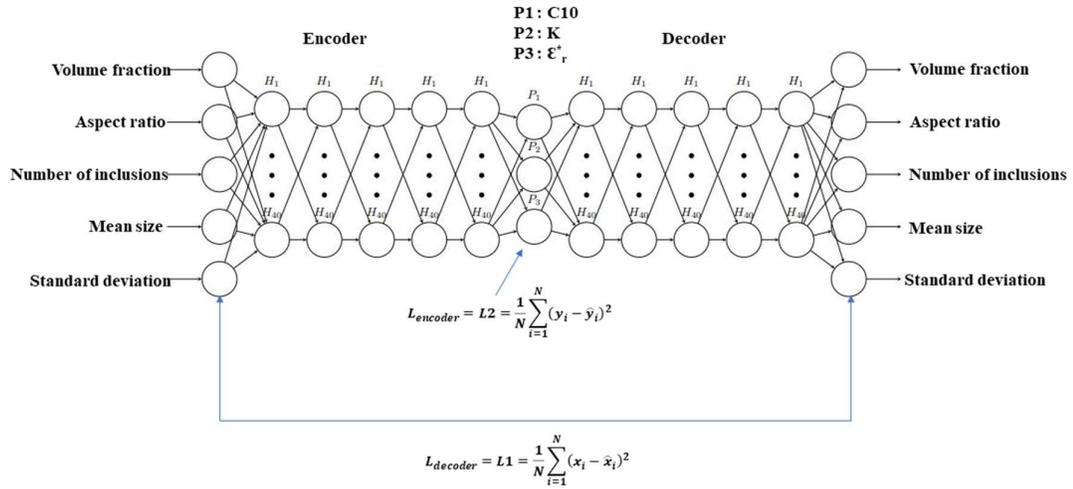

**Fig. 4.** Chosen VAE architecture through a grid-search consisting of its encoder functioning as the surrogate model of direct numerical homogenizers and its decoder acting as the inverse material designer in the inference mode ($H_i$ indicates the hidden neuron $i$ in the hidden layer).

The structure-property mapping between physical descriptors and material properties is unique in the built framework. This is because the uncertainty in the data generation and ML model is not considered in the current framework. The VAE model built is deterministic in nature and for each input/output, it gives one output/input after training when the weight parameters are fixed.

**Table 2:** The network architecture explained in detail

| ENCODER | | | DECODER | | |
|---|---|---|---|---|---|
| Layer | #Neurons | Activation function | Layer | #Neurons | Activation function |
| Input Layer | 5 | | Input Layer | 3 | |
| Hidden Layer #1 | 40 | | Hidden Layer #1 | 40 | |
| Hidden Layer #2 | 40 | ReLU | Hidden Layer #2 | 40 | ReLU |

| Hidden Layer #3 | 40 | | Hidden Layer #3 | 40 | |
| Hidden Layer #4 | 40 | | Hidden Layer #4 | 40 | |
| Hidden Layer #5 | 40 | | Hidden Layer #5 | 40 | |
| Output Layer | 3 | | Output Layer | 5 | |

A supervised machine learning model could be used for linking material properties and physical descriptors. Such a model can accurately predict the material properties from the physical descriptors. But in the inverse prediction, the accuracy reduces because of the loss of information of the inherent forward process. But the computational time is lower compared to the VAE. To prevent such reduction in accuracy due to loss of information, latent variables are used to capture this information in INNs. Even though INNs are accurate and computationally efficient, VAE is still preferred in our design. This is because there are multiple microstructures that give the same material properties which the INN cannot predict but the VAE can. The INN via the inverse design can predict the same microstructure that is used as the input but the VAE can predict a different microstructure as well. Hence, it was decided to choose a VAE here instead of INN.

3. **Results and discussions**

*3.1. Why our ML model is preferred to analytical models based on mean-field theories and experiments*

The hyperelastic parameters, $C_{10}$s, were predicted using the FE-optimization and an analytical method described in [34]. Since the numerical methods tend to be more accurate due to fewer simplifying assumptions and more details, the absolute difference between the values obtained by these two methods were divided by the numerical method results as the ground truths to achieve the error percentages of Fig. 5. The number of packs showing high discrepancy are quite high, pointing to the importance of high-fidelity numerical solutions for accurate modeling of the material response and its effective properties. The need for high fidelity solutions combined with their higher computational cost justifies the ML-based frameworks, such as the one developed in this study, to discover novel multi-functional composites and to predict their properties accurately and quickly. It should be noted that since the shear modulus of the matrix is very low, the non-linearity captured by the Neo-Hookean material model is low.

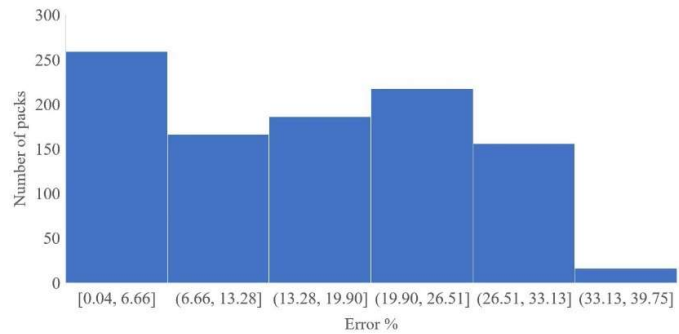

**Fig. 5.** Frequency histogram of the error percentage in calculation of the hyperelastic parameter based on the simple analytical model [34] with respect to the FE-optimization method.

Furthermore, the ML model predicts the composite dielectric constants and its thermal conductivity close to the experimental values obtained by Malakooti et al. [19] and Bartlett et all [15] as seen in Fig. 6(a) and Fig. 6(b), respectively. This is another indication that the model predicts accurate results for different possible physical microstructural descriptors while it can interpolate as well.

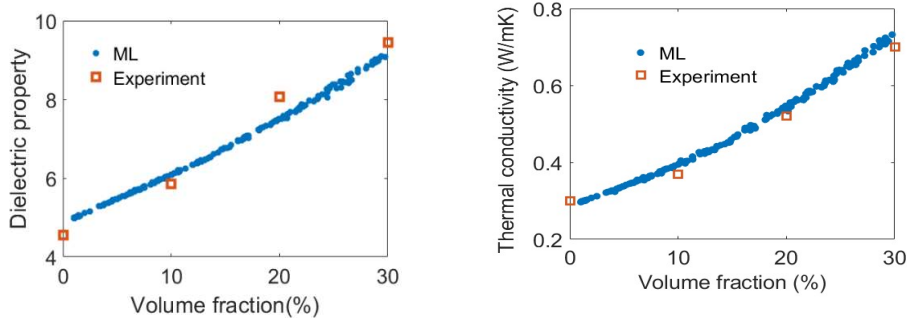

(a) (b)

**Fig. 6**. The ML predictions of (a) the composite dielectric constant and (b) its thermal conductivity for different volume fractions. The experimental data points are marked by squares for comparison.

*3.2. Distribution of material properties*

The distributions of material properties are shown in Fig. 7. Even though the microstructures are uniformly distributed according to their volume fraction, it can be clearly seen that the distribution of the material properties are random. Hence it can be concluded that the material properties do not depend solely on the volume fraction but also have dependency on other physical descriptors of the microstructure.

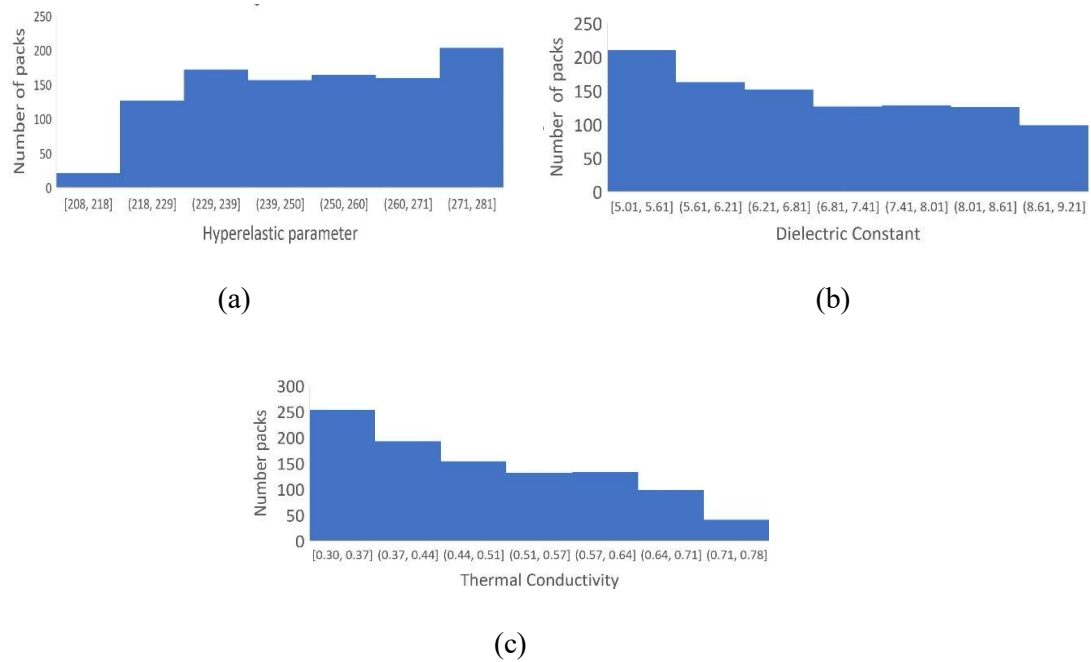

(a) (b)

(c)

**Fig.7**. Distribution of material properties (a) mechanical, (b) dielectric and (c) thermal properties

*3.3. VAE Performance as surrogate homogenizer and inverse material designer*

The FE and FFT simulations as well as the ML models were run and trained, respectively, on a workstation with a 16-Core AMD Ryzen 1950x processor and 32 GB DDR4 RAM, an HPC node with four 16-Core Intel 6130 processors and 3TB RAM, and a workstation with an Intel i7-11370H processor and 16GB RAM, respectively. The surrogate model is orders of magnitude faster than numerical simulations in predicting material properties, and the decoder is almost instantaneous in inverse designing, as shown in Table 2.

**Table 2:** Computational time comparison between numerical simulations and ML methods

| Mechanical simulations via FEA and optimization | ~2700 seconds/microstructure |
| --- | --- |
| Thermal simulations via FFT | ~1800 seconds/microstructure |
| Dielectric simulations via FFT | ~900 seconds/microstructure |
| Training time of the VAE model | ~ 540 seconds for training set |
| Inference time of surrogate model | ~0.00095 seconds for test set |
| Inference time of inverse model | ~0.00099 seconds for test set |

**Table 3:** Accuracy values calculated from different metrics

| | | Mean Maximum Error | Normalized RMSE | $R^2$ |
| --- | --- | --- | --- | --- |
| Dielectric Property | Surrogate | 0.034 | 0.012 | 0.998 |
| | Inverse | 0.088 | 0.036 | 0.984 |
| Thermal Property | Surrogate | 0.005 | 0.019 | 0.995 |
| | Inverse | 0.008 | 0.033 | 0.986 |
| Mechanical Property | Surrogate | 2.012 | 0.043 | 0.973 |
| | Inverse | 1.585 | 0.043 | 0.972 |

The accuracy values obtained for each material property using three different metrics are provided in Table 3. The mean maximum error and normalized RMSE are very low for thermal and dielectric properties indicating high accuracy. They are moderately low for mechanical property. The coefficient of determination of dielectric and thermal property is close to 1 and this indicates very high accuracy. To assess the accuracy of the surrogate model, the regression plots of the ML-predicted properties against the actual values of the properties obtained from the combined FE-optimization method and FFT are provided. It can be seen in Fig. 7(a) that the predicted dielectric constants are quite accurate ($R^2 = 0.998$) compared to those obtained from FFT simulations throughout the whole range of the volume fraction. From Fig. 7(b), the thermal conductivity values predicted by the model are quite accurate for low volume fractions. For higher volume fractions, there are very slight differences between the predictions and the true values, but the accuracy is still quite high ($R^2 = 0.995$). The predicted hyperelastic parameter is highly accurate for low volume fractions. The overall accuracy is moderately high, as can be seen in Fig. 7(c). The higher discrepancies in higher volume fraction packs may be due to the higher number of inclusions making high-concentration regions of LM inclusions. Increasing the number of microstructures with higher volume fractions in the dataset can increase the accuracy in predicting the mechanical properties. Even though the data is uniformly distributed, in the high volume fraction range for mechanical properties, the dependency between property and physical descriptors becomes nonlinear and hence a larger dataset is necessary to capture this nonlinear relationship with higher accuracy. Also, mechanical properties are not uniformly distributed as the dependency is not just on the volume fraction as seen in Fig. The artificially generated microstructures used in this study have a correlation between the concentration of particles and the volume fraction. Therefore, for high volume fractions, the LM inclusions are closer to each other than for lower volume fraction. This leads to more deviation from traditional mean field homogenization theories since the underlying assumptions in these theories is a uniform far-field stimuli [35,36] which is not the case in situations where inclusions are close to each other.

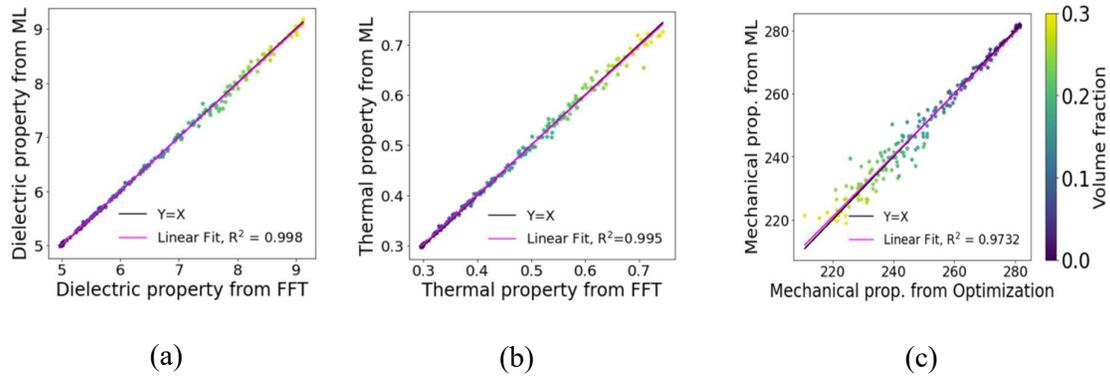

**Fig. 7.** Properties predicted by the ML surrogate model vs. those obtained from numerical simulations (the color bar represents volume fraction of inclusion) for (a) the dielectric constant, (b) the thermal conductivity, and (c) the hyperelastic parameter.

The predicted physical descriptors obtained from the decoder are verified with the help of the surrogate model that has been developed. They were used as inputs to the surrogate model to find their properties and to compare with the properties obtained from FE-Optimization and FFT homogenizations. According to Fig. 8(a), (b), and (c), the design predictions are quite accurate for most data points though the discrepancy for some data points is high. However, the accuracy of the decoder for inverse material designing is high $R^2 \sim 0.98$ for the dielectric property and the thermal property but is relatively low $R^2 \sim 0.97$ for the mechanical property. As mentioned earlier, it can be improved by augmenting the training dataset in regions with higher differences.

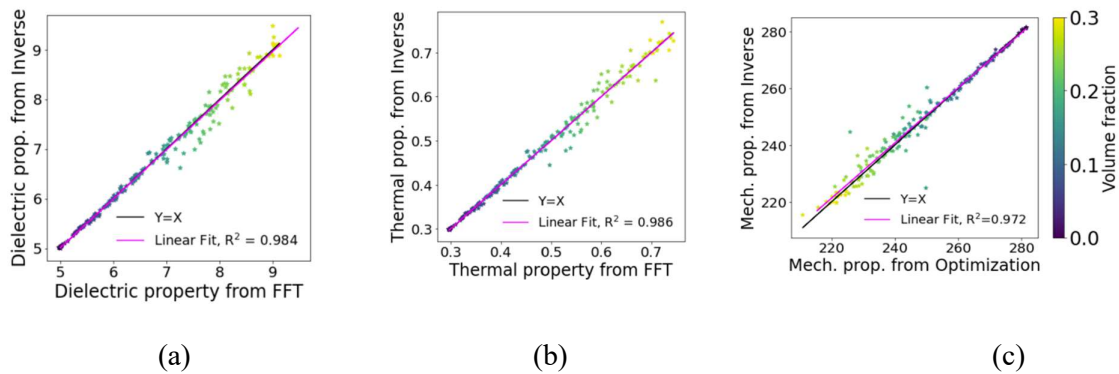

(a) (b) (c)

**Fig. 8.** Comparison of (a) dielectric, (b) thermal, and (c) mechanical properties obtained from numerical homogenizations with those obtained by plugging the physical descriptors predicted by the inverse model in the surrogate model.

From Fig. 7 and 8, it can be concluded that physical descriptors are sufficient to be used as input data instead of using microstructural images since accurate results are obtained just by using physical descriptors as inputs. Further, from Fig. 6, this can be concluded as the results also match well with experiments.

Further, even though there might be an argument that there could be loss of data if microstructural images are not used as inputs to the VAE, we can observe from Fig. 7 and Fig. 8 that the agreement of the predicted results is good with FFT and FEA results. Also, from Fig. 6 it can be concluded that the results also agree well with experiments. Hence it can be concluded from the results that there is no loss of data. In experiments, physical descriptors are used instead of images to synthesize composites. In case of particulate composites, micromechanical models consider physical descriptors and not images of microstructures. Using images instead of descriptors only leads to an increase in computational time. By the use of physical descriptors, the ML model is simplified and the computational time is reduced and at the same time, accurate results are predicted.

*3.4. Relationship between the physical descriptors and material properties*

The material property surfaces as functions of microstructral descriptors are visualized in Fig. 9 by each property's contour plots. The 2D grids of the microstructral descriptors were considered as inputs of the trained surrogate model to map each point to its corresponding properties. Fig. 9(a) and (d) indicate that the dielectric constant is heavily and nonlinearly influenced by the volume fraction while it has small linear dependence on the aspect ratio and the mean particle size. Also, some irregularities can be seen in regions with very high volume fractions as there were fewer available data points in the associated DoE regions which negatively affect the ML accuracy. Similar trends can be inferred for the composite thermal conductivity (Fig. 9(c) and (f)) as it is also a transport porperty following governing equations similar to those of electrostatics used for the dielectric constant homogenization. According to mechanical Fig. 9(b) and (e), there is no regular trend except that higher volume fractions lead to higher hyperelastic parameter. This further shows the nontrivial nature of mechanical homogenization of the composites with the liquid metal inclusions as discussed in section 3.1. Again, further investigation is necessary for high volume fractions with high number of particles. Finally, it can be concluded that the relationship between the material properties and microstructural descriptors is nonlinear and complex in general, and the ML surrogate model enables us to infer such relationships in the studied material system.

The dependence on volume fraction in our study is similar to that seen in micromechanical based mean field homogenization models seen in [35,36]. The various models considered in these studies are the Mori-Tanaka method, the Eshelby method, the double inclusion model and the differential method. The general trend with these homogenization models is the increase in material properties with volume fraction and the same is observed in our results from the VAE. This is because the properties are dependent on the concentration of the LM inclusions in the matrix. Thermal conductivity is determined in [35] for the same material system using the Mori-Tanaka method for different aspect ratios. The dependency on aspect ratios is less at low volume fractions and our results agree with this conclusion. Further, the double inclusion model is used in [36] to predict the thermal conductivity and dielectric constant for a similar material system with the same inclusion material and a different matrix material for various inclusion sizes. This prediction shows that there is almost no dependency on the size at low volume fractions for the dielectric property and the same results are obtained in our studies. The thermal conductivity increases slightly with size over 20% volume fraction. We can see a slight increase in our results studies as well. Thus, it can be concluded that our results agree with micromechanical models.

The micromechanical models used in [35] underpredict thermal conductivities as compared to experimental results in the volume fraction regime used in our studies. But the results predicted by the VAE is closer to experimental results and hence it can be concluded that the VAE is more accurate than the micromechanical mean field homogenization models.

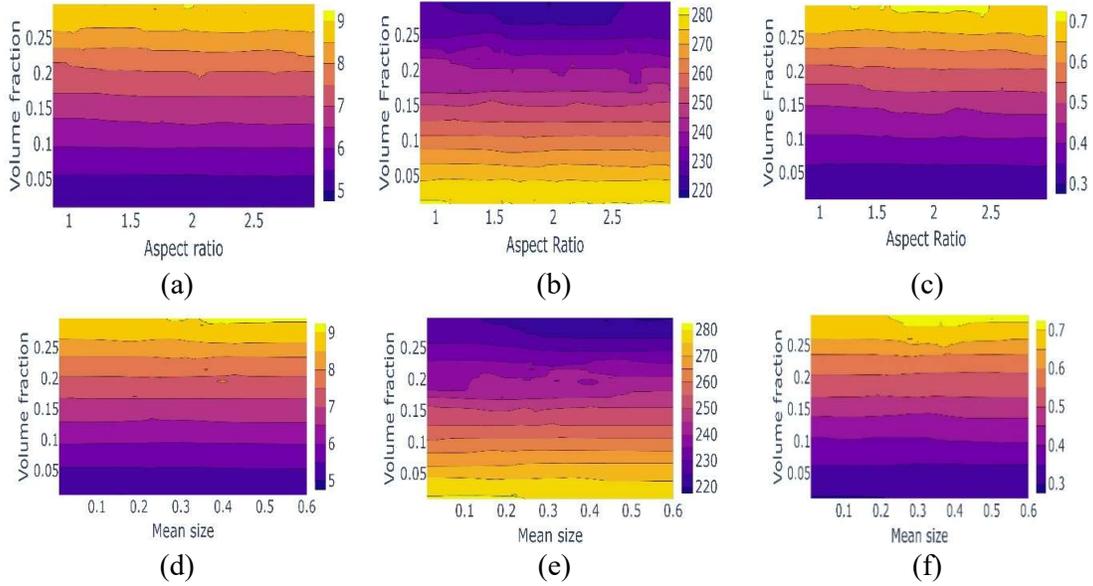

**Fig. 9.** The contour plots of composite properties with respect to different physical microstructural descriptors: the first column for the dielectric property, the second one for the mechanical property, and the third one for the thermal property.

4. **Conclusions**

Our study envisions a general computational framework based on a semi-supervised ML method to link particulate composite microstructures to their multifunctional properties given appropriate training datasets. Our VAE's encoder can predict the material properties for a given set of physical microstructural descriptors instead of complex and time consuming direct numerical solutions of homogenization problems such as FE and FFT, while its decoder can design material microstructures based on the target properties. In the process of data generation, we found out that our FE-optimization method is much more accurate in the prediction of effective or homogenized hyperelastic parameter than that of continuum micromechanics especially for packs with a high number of inclusions concentrated near each other. Due to the packing algorithm limitation, the generated dataset is limited in terms of the volume fractions (up to 30% for LM inclusions). Thus, future studies may explore higher volume fractions with higher number of particles to better understand and utilize the studied material system. Another valuable byproduct of our study is a dataset of orthotropic microstructures, following the unidirectional, tensile, FE simulations, similar to the ones synthesized by experimentalists. This has great potential in guiding the researchers in functionalizing LMEEs in different applications by tuning their properties in the principal stretch directions.